# An analysis of citing and referencing habits across all scholarly disciplines: approaches and trends in bibliographic referencing and citing practices

Erika Alves dos Santos[a], Silvio Peroni[b], Marcos Luiz Mucheroni[c]

[a] School of Communication and Arts (ECA), Department of Information & Culture (CBD), University of São Paulo (USP), São Paulo, SP – Brazil; Digital Humanities Advanced Research Centre (DHARC), Department of Classical Philology and Italian Studies, University of Bologna, Bologna, Italy; Fundação Jorge Duprat Figueiredo de Segurança e Medicina do Trabalho (Fundacentro), São Paulo, SP – Brazil, erika.santos@fundacentro.gov.br

[b] Research Centre for Open Scholarly Metadata, Department of Classical Philology and Italian Studies, University of Bologna, Bologna, Italy; Digital Humanities Advanced Research Centre (DHARC), Department of Classical Philology and Italian Studies, University of Bologna, Bologna, Italy, silvio.peroni@unibo.it

[c] School of Communication and Arts (ECA), Department of Information & Culture (CBD), University of São Paulo (USP), São Paulo, SP – Brazil, mucheroni.marcosl@gmail.com

## Abstract

**Purpose.** In this study, we want to identify current possible causes for citing and referencing errors in scholarly literature to compare if something changed from the snapshot provided Sweetland in his 1989 paper.

**Design/methodology/approach.** We analysed reference elements, i.e. bibliographic references, mentions, quotations, and respective in-text reference pointers, from 729 articles published in 147 journals across the 27 subject areas.

**Findings.** The outcomes of our analysis pointed out that bibliographic errors have been perpetuated for decades and that their possible causes have increased, despite the encouraged use of technological facilities, i.e., the reference managers.

**Originality.** As far as we know, our study is the best recent available analysis of errors in referencing and citing practices in the literature since Sweetland (1989).

# 1. Introduction

Contextualising and supporting the tracking of historical approaches is one of the functions played by citations (Karcher, & Zumstein, 2018). As regulation and mediation instruments between citations and readers, reference styles directly influence how we cite and read and indirectly how we follow back on the author's thinking and research. Appropriate use of reference styles grants the accomplishment of these purposes by providing clear guidelines under which cited works should be formatted to be correctly retrieved.

A study by Sweetland (1989) highlighted the functions of bibliographic references and style manuals and, in particular, the errors in the reference lists and in-text citations that represent a crucial issue for accomplishing the citations' functions. Sweetland's findings pointed out that the great variety of formats for referencing cited articles, added to the lack of agreement among journals or authors, increases the chances of misunderstanding referencing guidelines which, consequently, contributes to the high rates of errors in bibliographic metadata description. The study also pointed out other causes for bibliographic errors: the lack of commitment of publishers to the normalisation of citation metadata, the diffusion of responsibility in the publishing process, the lack of training in the norms and purposes of the bibliographic citation, the misleading of citation rules (i.e., the reference styles' contents), the misunderstanding of foreign languages, the human inabilities to correctly reproduce long information strings and, the failure to examine the documents cited.

Since changes in information representation are still in progress, this article expands and updates Sweetland's investigation and some initial analysis we conducted in some specific subject areas – i.e. Medicine and Social Sciences (Santos et al., 2021a) – and in the identification of the standard metadata used in bibliographic references depending on the publication types (Santos et al., 2023), by discussing the role played by citing and referencing metadata, reference styles, and their trends, by answering two research questions (RQ1-2):

RQ 1 - Considering the changes in the production, storage, retrieval, and use in the information universe, do Sweetland's (1989) claims remain applicable?

RQ 2 - Are there current possible causes for citing and referencing errors other than those pointed out by Sweetland's study?

Reflections on these issues are aligned to Information Science's accomplishment of its mission, which goes back to the Five Laws of the Library Science proposed by the Indian mathematician and library science scholar Sr. Shiyali Ramainrita Ranganathan (1892–1972) in 1931. Referencing and citing issues are complex, especially considering the current descriptive representation's revision context. Referencing and citing are vast and multifaceted activities. Because of this, this article is not intended to deal fully with all aspects involved in these matters. It should, therefore, be taken as a starting point for a significant discussion on ways of becoming citing and referencing tasks less laborious and time-consuming.

## 2. Material and methods

Data supporting this research were extracted from the SCImago Journal & Country Rank, "an open access scientometric directory" (Guerrero-Bote & Moya-Anegón, 2012, p. 675) that includes the journals and country scientific indicators developed from the information contained in the Scopus database. SCImago covers the 27 major thematic areas divided into 313 specific subject categories (also referred to as disciplines in this study), comprising over 34.000 journal titles from more than 5.000 international publishers and country performance metrics from 239 countries worldwide.

The selection of journals and their related articles for composing our study's sample considered the most cited journals in each thematic area in the 2015-2017 triennium according to the SCImago total citation ranking[1]. On a smaller scale, the sample reproduced the proportional representation of each thematic area at SCImago Ranking in terms of dimension. In particular, we established the representativeness of each subject area in SCImago by calculating the percentage of the total number of journal titles indexed under that subject area against the total number of journals indexed in the SCImago. Ideally, the more a subject area was represented in SCImago, the more was the number of journals to select for such a subject area. Anyway, the minimum number of journals admitted for each subject area was 2, regardless of the percentage representativeness of the subject area. In addition, we admitted only one journal from each publisher under the same subject area, avoiding results based on the same editorial policy. The analysis assumed the data contained in SCImago Ranking as shown within the database in the period covered by the data gathering[2].

---

[1] As explained in (Santos et al., 2020), we used the most cited metrics for journals as a proxy for measuring the prestige of journals, assuming that the more citations one journal receives, the more relevant it is supposed to be in its particular subject area.

[2] For a more detailed description of the methodological procedures of this study, please consult the supplemental protocol available in (Santos et al., 2020).

The sample took the date of 31 October 2019 as an upper-bound reference from which the most recent journal issues fully attending the selection criteria were selected. Articles published between 1 October 2019 and 31 October 2019 (or the most recent equivalent period for journals not publishing articles in this interval) were considered for selecting the journals in case they did not organise their articles to any issue. Special issues were not regarded as eligible for composing the sample since the editorial policies with which they are submitted may differ from those applied to regular issues. For the same reason, articles that were not original research communications, like special articles, letters to the editor, and book reviews, were not considered eligible for the composition of our sample.

SCImago's thematic scheme is based on The All Science Journal Classification (ASJC), developed by Elsevier and adopted by the Scopus database, which was considered by this analysis for grouping journals according to thematic categories and subcategories to support approaching related disciplines journals data.

Each journal is represented by five articles published in the most recent issue between October $1^{st}$ and October $31^{st}$, 2019. For journals not releasing any issue in this period, the sample considered the immediately previous issue published before October $1^{st}$. For issues containing more than five articles, the selection considered a probabilistic systematic random sampling technique based on the average number of articles published by the journal in the period above. As for the journals containing fewer than five articles, the sample considered all those attending the selection criteria.

The gathering of restricted access articles was guaranteed by using either the journal subscription programme at the University of Bologna and the University of São Paulo or other available browser plugins for finding open-access versions of the articles behind the paywall. The articles were gathered using version 79 of Google Chrome browser and an Italian IP with a personal Internet connection.

The analysis addressed mentions (i.e. rephrasing a passage or idea introduced in a cited work without quoting it explicitly), quotations (i.e. a reference to an explicit textual passage of a cited work reported in the citing work), their respective in-text reference pointers (i.e. the textual devices, such as "[3]", that denote the bibliographic references related to mentions and quotations), and bibliographic references. All these items were extracted manually by us from the journal articles and were analysed both from qualitative and quantitative perspectives to demonstrate the dynamic relationship between them, the interdependence in the fulfilment of their primary purposes (and their relation with bibliographic catalogues), and the importance on assuring "clear, precisely, stated, and commonly shared understanding" metadata within all formats of representing information comprised by descriptive representation (IFLA Study Group on the Functional Requirements for Bibliographic Records, 2009, p. 2).

## 3. Data

The sample comprises 729 articles (we retrieved in PDF format) published in 147 journals from 27 subject areas, arranged under four main top subject categories, as shown in Table 1.

| Subject Categories | Subject Areas | Total number of journals | Total number of articles | Freely available journals | Freely available articles | Avg number of authors per article | Total bibliographic references | Total number of mentions | Avg mentions per article | Total number of quotations | Avg quotations per article |
|---|---|---|---|---|---|---|---|---|---|---|---|
| Health Sciences | Medicine | 27 | 132 | 33% | 42% | 9 | 5,340 | 8,400 | 64 | 5 | 0.04 |
| | Nursing | 2 | 10 | 0% | 50% | 2 | 440 | 768 | 77 | 17 | 1.70 |
| | Veterinary | 2 | 10 | 0% | 50% | 5 | 336 | 561 | 56 | 0 | 0.00 |
| | Dentistry | 2 | 10 | 0% | 10% | 6 | 381 | 533 | 53 | 0 | 0.00 |
| | Health Professions | 2 | 10 | 50% | 50% | 4 | 271 | 446 | 45 | 1 | 0.10 |
| | Health Sciences overall rates | Total 35 | Total 172 | Avg 16.6% | Avg 40.4% | Avg 5.2 | Total 6,768 | Total 10,708 | Avg 59 | Total 23 | Avg 0.37 |
| Social Sciences | Arts and Humanities | 10 | 45 | 20% | 40% | 3 | 1,752 | 2,556 | 57 | 614 | 13.64 |
| | Business, Management and Accounting | 5 | 25 | 40% | 40% | 2 | 1,642 | 2,719 | 109 | 73 | 2.92 |
| | Decision Sciences | 2 | 10 | 0% | 50% | 2 | 428 | 749 | 75 | 0 | 0.00 |
| | Economics, Econometrics and Finance | 2 | 10 | 0% | 75% | 2 | 457 | 610 | 61 | 1 | 0.10 |
| | Psychology | 4 | 20 | 0% | 5% | 4 | 1,536 | 2,959 | 148 | 32 | 1.60 |
| | Social Sciences | 17 | 81 | 35% | 52% | 2 | 4,571 | 6,953 | 86 | 835 | 10.31 |
| | Social Sciences overall rates | Total 40 | Total 191 | Avg 15.8% | Avg 43.6% | Avg 2.5 | Total 10,386 | Total 16,546 | Avg 89.33 | Total 1555 | Avg 4.76 |
| Life Sciences | Agricultural and Biological Sciences | 8 | 40 | 25% | 40% | 5 | 2,281 | 3,445 | 86 | 1 | 0.03 |
| | Biochemistry, Genetics and Molecular Biology | 8 | 39 | 25% | 41% | 6 | 2,229 | 3,524 | 90 | 0 | 0.00 |
| | Immunology and Microbiology | 2 | 10 | 0% | 50% | 6 | 521 | 786 | 79 | 0 | 0.00 |
| | Neuroscience | 2 | 10 | 50% | 80% | 6 | 606 | 848 | 85 | 0 | 0.00 |
| | Pharmacology, Toxicology and Pharmaceutics | 3 | 15 | 33.3% | 53,3% | 9 | 821 | 1,213 | 81 | 0 | 0.00 |
| | Life Sciences overall rates | Total 23 | Total 114 | Avg 26.6% | Avg 52.8% | Avg 6.4 | Total 6,458 | Total 9,816 | Avg 84.2 | Total 1 | Avg 0.01 |
| Physical Sciences | Chemical Engineering | 2 | 10 | 50% | 50% | 5 | 452 | 706 | 71 | 0 | 0.00 |
| | Chemistry | 3 | 15 | 33.3% | 40% | 6 | 682 | 846 | 56 | 0 | 0.00 |
| | Computer Science | 8 | 39 | 0% | 20% | 3 | 1,125 | 2,167 | 56 | 0 | 0.00 |
| | Earth and Planetary Sciences | 4 | 20 | 25% | 50% | 7 | 1,157 | 1,942 | 97 | 0 | 0.00 |
| | Energy | 2 | 10 | 0% | 10% | 3 | 583 | 924 | 92 | 0 | 0.00 |
| | Engineering | 10 | 48 | 30% | 35.4% | 4 | 1,470 | 2,148 | 45 | 13 | 0.27 |
| | Environmental Science | 5 | 25 | 20% | 32% | 4 | 1,547 | 2,147 | 86 | 43 | 1.72 |
| | Materials Science | 4 | 20 | 0% | 5% | 5 | 845 | 1,156 | 58 | 1 | 0.05 |
| | Mathematics | 5 | 25 | 20% | 60% | 3 | 569 | 900 | 36 | 0 | 0.00 |
| | Physics and Astronomy | 4 | 20 | 75% | 50% | 4 | 1,101 | 1,758 | 88 | 2 | 0.10 |
| | Physical Sciences overall rates | Total 47 | Total 232 | Avg 25.3% | Avg 35.24% | Avg 4.4 | Total 9,531 | Total 14,694 | Avg 68.5 | Total 59 | Avg 0.21 |
| Multidisciplinary | Multidisciplinary | 4 | 20 | 25% | 35% | 11 | 997 | 1,697 | 85 | 1 | 0.05 |
| | Multidisciplinary overall rates | Total 4 | Total 20 | Avg 25% | Avg 35% | Avg 11 | Total 997 | Total 1,697 | Avg 85 | Total 1 | Avg 0.05 |
| Total 5 subject categories | Total sample overall rates | Total 149 | Total 729 | Avg 21.8% | Avg 41.4% | Avg 4.74 | Total 34,140 | Total 53,461 | Avg 75 | Total 1,639 | Avg 2 |

**Table 1. Journals sample quantitative overview ("Avg" means average)**

On average, co-authorship was more frequent among Multidisciplinary articles compared with articles of the other subject categories. Data collected within the scope of this investigation did not demonstrate reasons that justify this behaviour and, therefore, may configure an object of further studies.

The sample size reproduces the representativeness of each subject area according to SCImago. All the results considered in this study were based on the data extracted from articles composing the sample and from information available from publishers' webpages as provided during data gathering, i.e., from November 2019 to May 2020.

Regarding publishers' nationality, 18 countries were represented within the journal sample. On average, the leading three publisher's nationalities per subject category are as follows: Health Sciences: United Kingdom (46%), United States (38%) and Japan (10%); Social Sciences: United States (40%), United Kingdom and Netherlands (both with 26%); Life Sciences: United States (32%), Switzerland (24%) and United Kingdom (20%); Physical Sciences: United States (33%), United Kingdom (23%) and Netherlands (20%) and lastly, Multidisciplinary with the United Kingdom and United States (both with 50%). This allows us to consider the United Kingdom, the United States, and the Netherlands as the most contributing countries in the journals sample, with a respective average representativity of 36.01%, 27.74%, and 16.62% of the whole sample.

100% of analysed journals provided their articles in downloadable PDF files. From this, 72.8% of journals also provided articles in HTML format. Regarding the modality of access, 40.4% of the articles composing the sample were freely readable, against 59,6% of restricted access articles.

All gathered raw data presented in the following sections are available in (Santos et al., 2021b).

## 4. Results

Life Sciences showed the highest citing habits rates with an average of 56 bibliographic references per article, closely followed by the Social Sciences subject category with an average of 54 bibliographic references per article. All the sample articles contain mentions, while only 19.2% contain quotations.

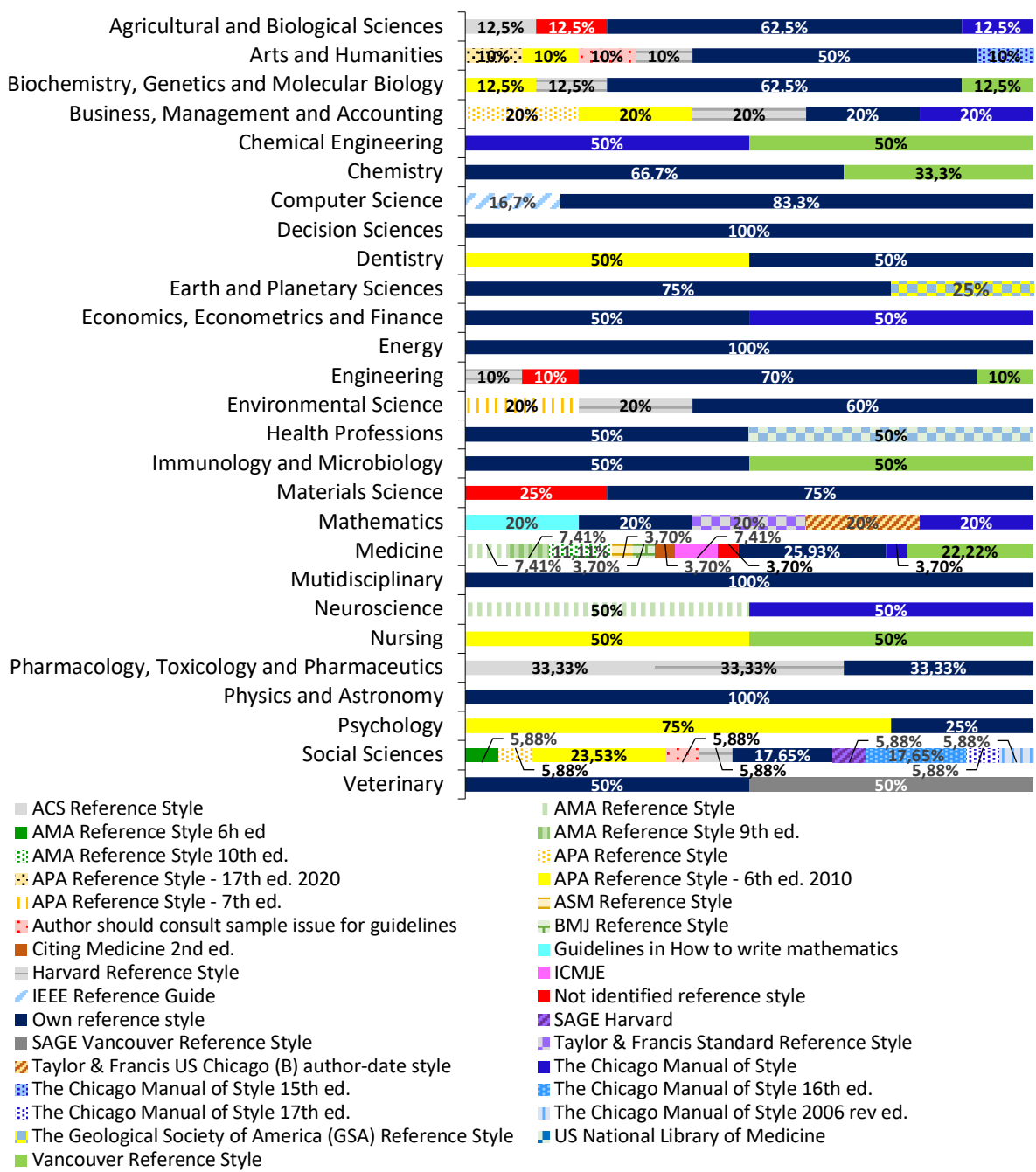

**Graphic 1. Percentual representation of reference styles adoption by subject area**

According to Graphic 1, the four most adopted reference styles are APA 6$^{th}$ ed. 2010 (8.05% of total journals sample), Vancouver (10.73% of entire journals sample), and Chicago (4.69% of total journals sample). Those referred to as "own reference styles" (46.97% of total articles sample) correspond to customised versions of widely accepted reference styles, i.e., Vancouver, Chicago, and APA or reference styles authored by the publishers. Among the standards reference styles, Vancouver proved to be the most adopted one across disciplines (10.73% of journals sample), followed by the APA Reference Style (6$^{th}$ ed. 2010), which achieved a rate of 8.05% of adoption across disciplines also showed to be the most adopted within Social Sciences' journals (52.94%). The most adopted reference style in Life Sciences journals was Chicago (no edition specified), with 8.69% adopting journals and 4.69% adopting journals across all subject categories.

In total, we found 31 guidelines on formatting citing and referencing data and metadata provided or indicated by publishers in their journal's webpages. Fifteen different reference styles were detected within journals composing the Social Sciences subject category, disregarding the multiple editions of a single reference style. Indeed, Social Sciences was the subject category with the most comprehensive range of reference styles, followed by Health Sciences (14 reference styles), Physical Sciences (11 reference styles), Life Sciences (8 reference styles), and lastly, the Multidisciplinary subject category in which only one reference style was detected (i.e., "own reference style").

Our sample showed the adoption of around three reference styles per subject area. Medicine and Social Sciences subject areas overcame the average with 11 and 10 different adopted reference styles, respectively. Arts and Humanities, Business, Management and Accounting, and Mathematics subject areas adopted five different reference styles each – all comprising journals adopting "own reference styles".

Only 48.14% of subject areas adopted two different reference styles: Chemical Engineering; Chemistry; Computer Sciences; Dentistry; Earth and Planetary Sciences; Economics, Econometrics and Finance; Health Professions; Immunology and Microbiology; Materials Science; Neuroscience; Nursing; Psychology, and Veterinary.

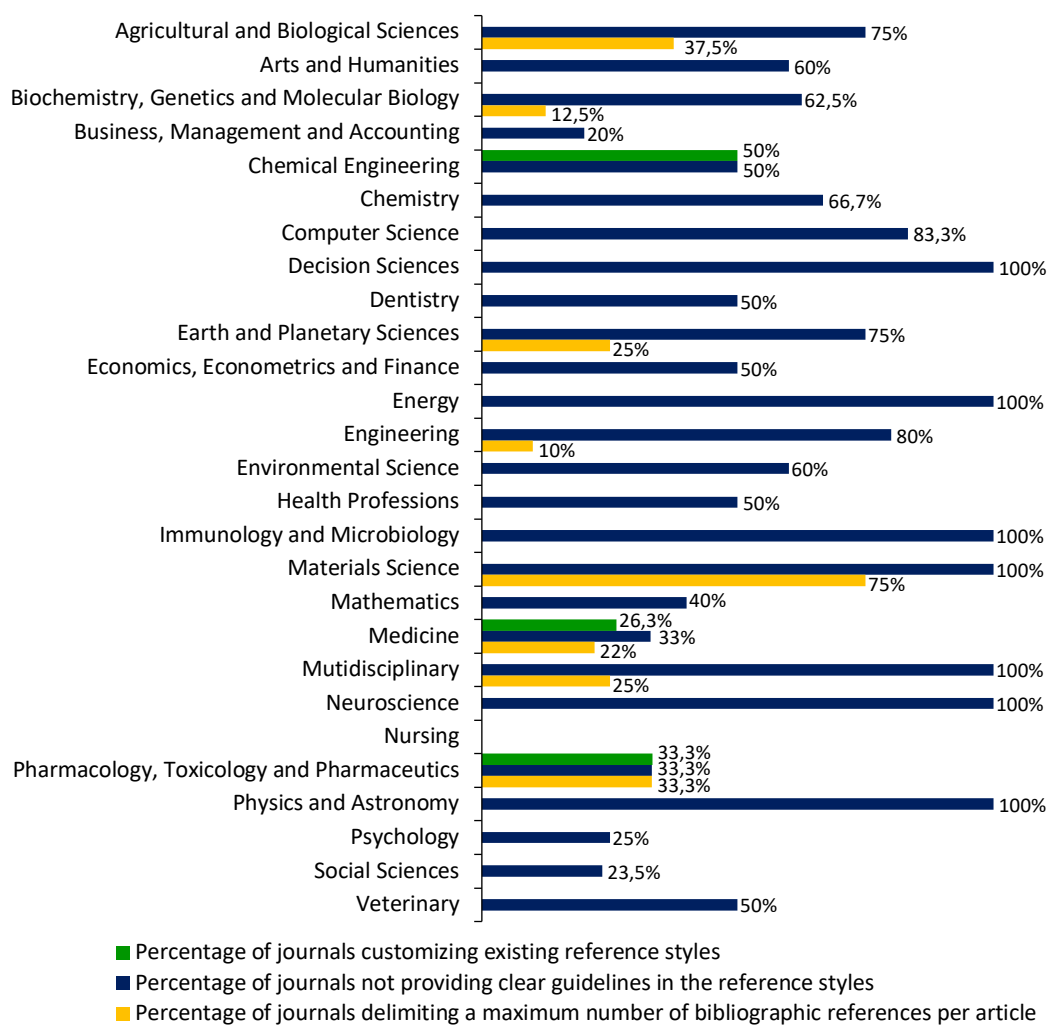

**Graphic 2. Percentual distribution of journals considering aspects from the respective reference styles**

Graphic 2 shows an analysis of journals adopting their "own reference styles", as introduced in Graphic 1. This analysis considered whether the reference styles provided enough instructions for metadata description, considering mainly the following aspects:

a) guidelines coverage – we evaluated whether the reference styles provide clear guidelines[3] for (at least) the most frequently cited types of publications, i.e., books, articles, proceedings, and correlated papers and events, electronic content available online, and grey literature;

---

[3] By "clear guidelines" we mean an extensive and comprehensive set of guidelines for describing and formatting citing and referencing metadata, i.e., in-text reference pointers concerning mentions and quotations, and bibliographic references, considering the articles analysed in this study. Thus, we will focus empirically on the actual references found (and their publication type), without considering potential material that may be cited but for which we did not find an occurrence in our document corpus. Journals clearly stating the adopted reference styles were considered as "clear guidelines providers", based on the understanding that by clearly stating the adopted reference styles, the responsibility of providing clear instructions on this matter automatically falls upon the reference style itself, instead of the publisher.

b) guidelines specificity – we considered the level of detail of the guidelines, i.e. if it provides information on how to structure basic metadata (such as authors, title, year of publication, venue, identifiers, etc.) necessary for a user to understand what kinds of publication is referenced.

62.49% of reference styles were classified as "not providers of clear guidelines" in our analysis. In 70% of the subject areas, the average rates of journals adopting reference styles classified as "not clear" are equal to or higher than 50%. In 18.51% of the sample (i.e., Decision Sciences, Energy, Materials Science, Physics and Astronomy and Multidisciplinary), the rate of journals adopting "own reference styles" classified as "not clear" is 100%. From the subject area perspective, some disciplines showed a critical scene in which 100% of the reference styles were addressed as "not clear", namely Decision Sciences (corresponding to 16.66% of the Social Sciences subject category); Immunology and Microbiology and Neuroscience (corresponding to 40% of Life Sciences subject category); Energy, Material Science, Physics and Astronomy (corresponding to 30% of Physical Sciences subject category) and Multidisciplinary subject area (corresponding to 100% of the Multidisciplinary subject category). The best behaviour concerning the clarity of the journal reference style adopted was observed within the Nursing subject area. No journal was classified as a provider of unclear guidelines.

On average, 54.96% of journals need to provide clear guidelines concerning citing and referencing, and 6.12% of publishers, on average, limit the maximum number of bibliographic references allowed per article. These limits usually corresponded to 30 bibliographic references, even if someone considered up to 50 bibliographic references in specific situations. The Multidisciplinary subject category was the one more affected by the constraint of the total number of bibliographic references per article (25%), followed by Life Sciences (17%) and Health Sciences and Physical Sciences (4% each). However, we noticed that some articles have more bibliographic references than the limit established by the publisher. Social Sciences, instead, was the most flexible subject category concerning the extent of bibliographic references, with no publisher setting a threshold. The only three disciplines in which the average number of bibliographic references per article was under the limit of 30 bibliographic references per article were Health Professions (27.10), Computer Science (28.85), and Mathematics (22.76). The averages of bibliographic references per article are addressed within Graphic 3.

---

Since the purposes of this work do not provide for the analysis of widely adopted reference styles' contents, i.e., Vancouver and APA, verifying the level of clarity of those guidelines remains an open question to be approached in further studies.

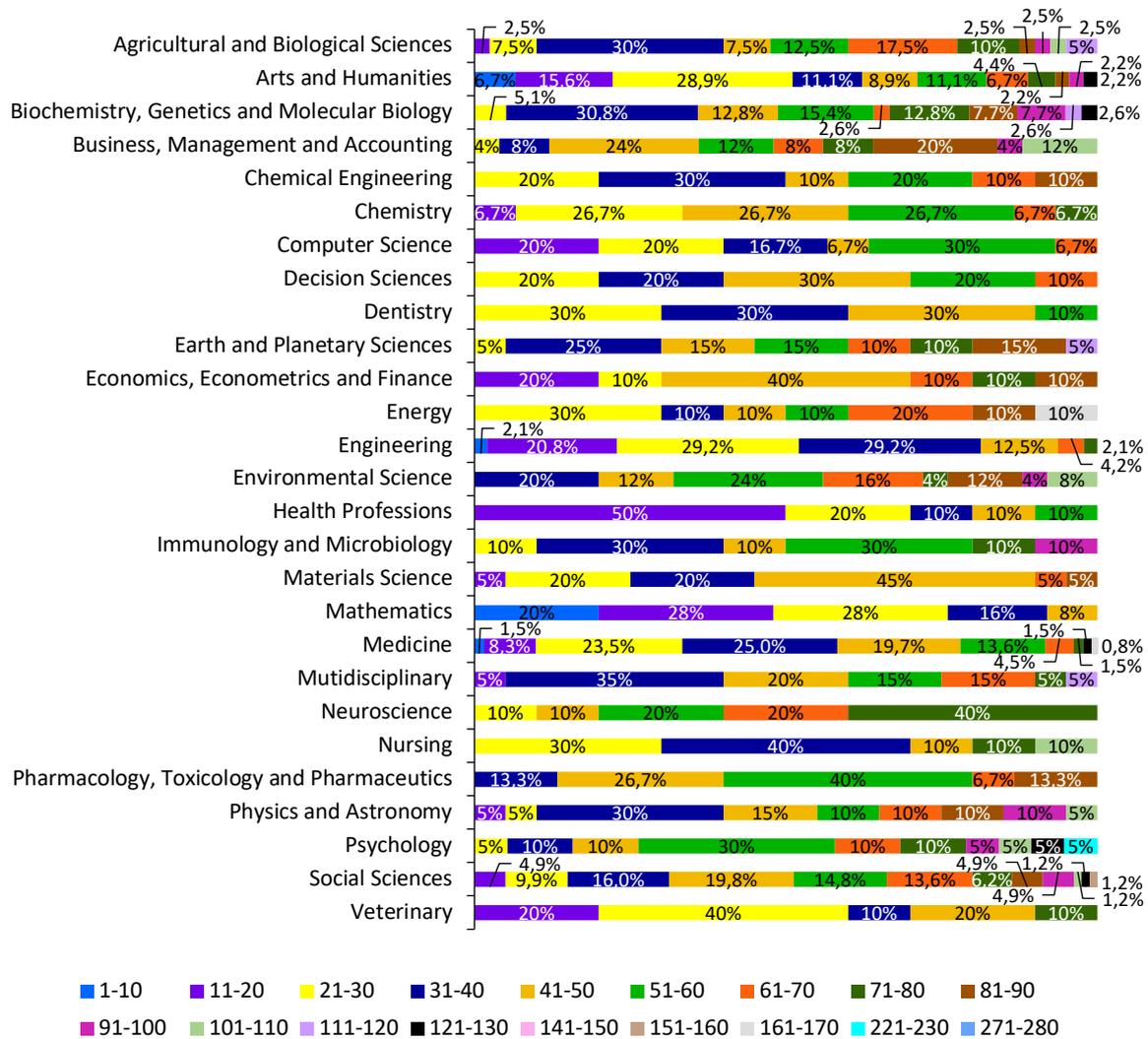

**Graphic 3.** Distribution of articles per subject area, according to the number of bibliographic references included in bibliographic references lists. It is worth mentioning that the values 131-140, 171-220 and 231-270 were not observed in the gathered data and, therefore, are not shown.

Our analysis identified two citation systems within articles composing the sample, according to data represented within Graphic 4, i.e., the citation-sequence system (which uses numbers to refer to bibliographic references within the text, e.g., "[3]") and the author-date system (which uses the names of some authors of the article plus a date, e.g., "(Doe et al., 2020)"). On average, the citation-sequence system is adopted by 51.37% of the article's sample, against 48.41% adopting the author-date system. This scenario prevents assuming one of the citation systems as the most adopted overall. However, by considering the subject categories perspective, we can say that Health Sciences, Physical Sciences, and Multidisciplinary journals tended to adopt citation-sequence system (respectively 79%, 57% and 51% of journals, on average), Social Sciences and Life Sciences tended to adopt author-date system (respectively 86% and 52% of journals, on average). It is worth mentioning that some articles adopted author-date and citation-sequence simultaneously in the Social Science subject area.

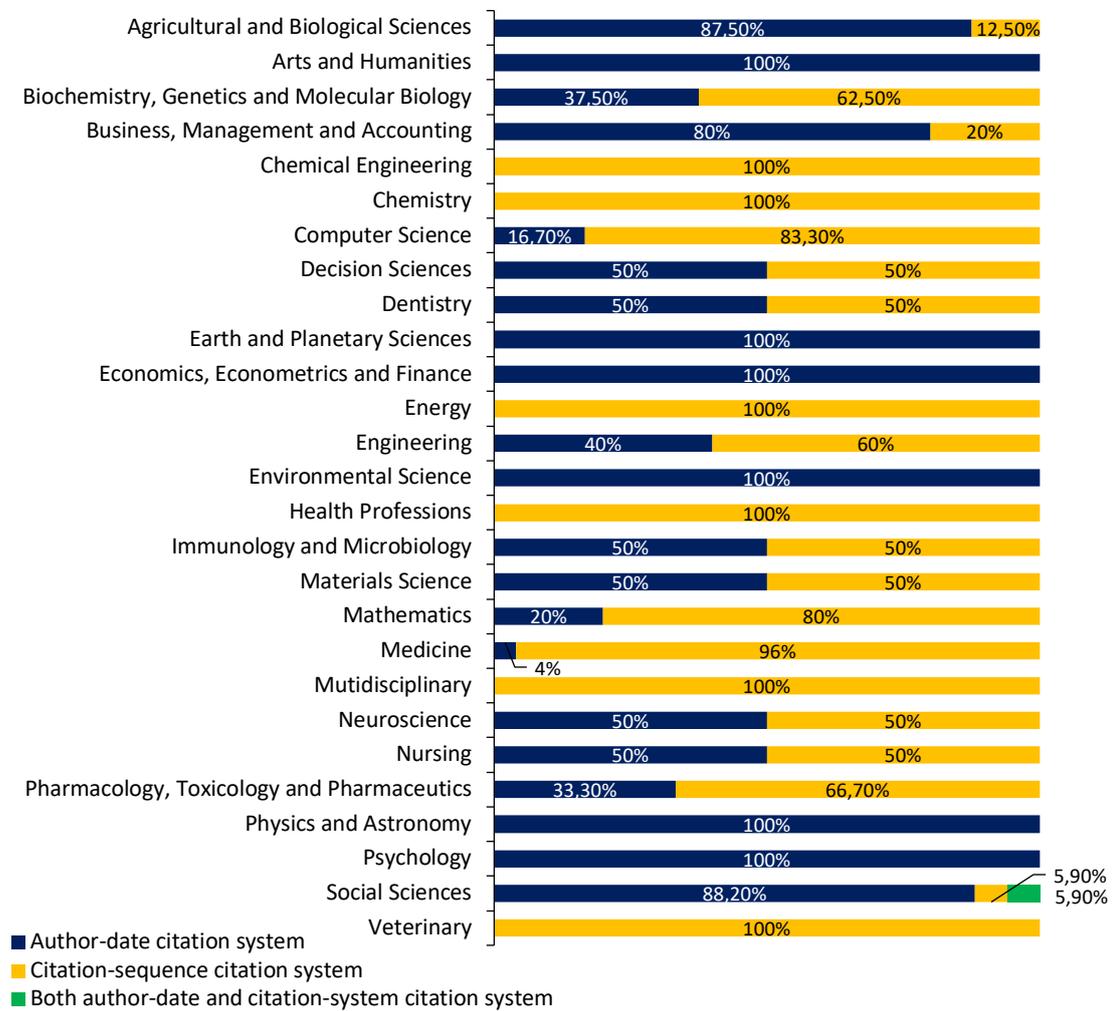

**Graphic 4. Percentual distribution of journals per subject area, according to the citation system adopted**

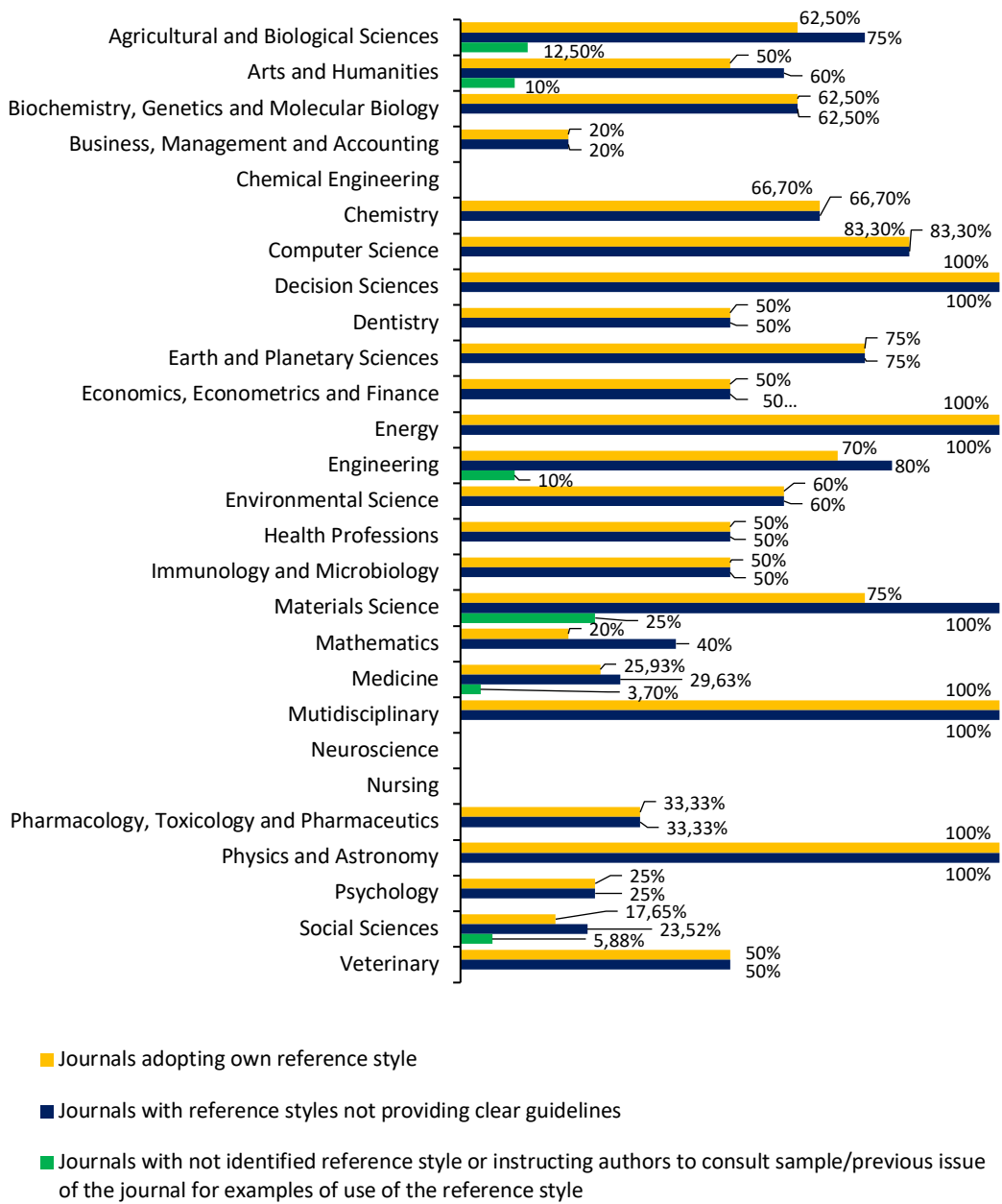

- Journals adopting own reference style
- Journals with reference styles not providing clear guidelines
- Journals with not identified reference style or instructing authors to consult sample/previous issue of the journal for examples of use of the reference style

**Graphic 5. Approaches concerning the relation between reference styles authored by publishers and the level of clarity of the guidelines for formatting citing and referencing data**

Graphic 5 summarises the correspondence between the rates of publishers adopting "own reference styles" and the rates of reference styles not providing precise referencing and citing instructions. The portion of journals for which it was not possible to identify the adopted reference style was relatively low (1.90% of journals on average). However, 10% of Arts and Humanities journals and 5.88% of Social Sciences journals did not clearly state the adopted reference style. They instructed authors to consult any previous journal issue or a sample article provided by the publisher to understand guidelines for citing and referencing. Although the elements provided by publishers are insufficient even to classify the instructions concerning their level of clarity, they were put together in this graphic to support the statements on how publishers may be negligent concerning citing and referencing matters.

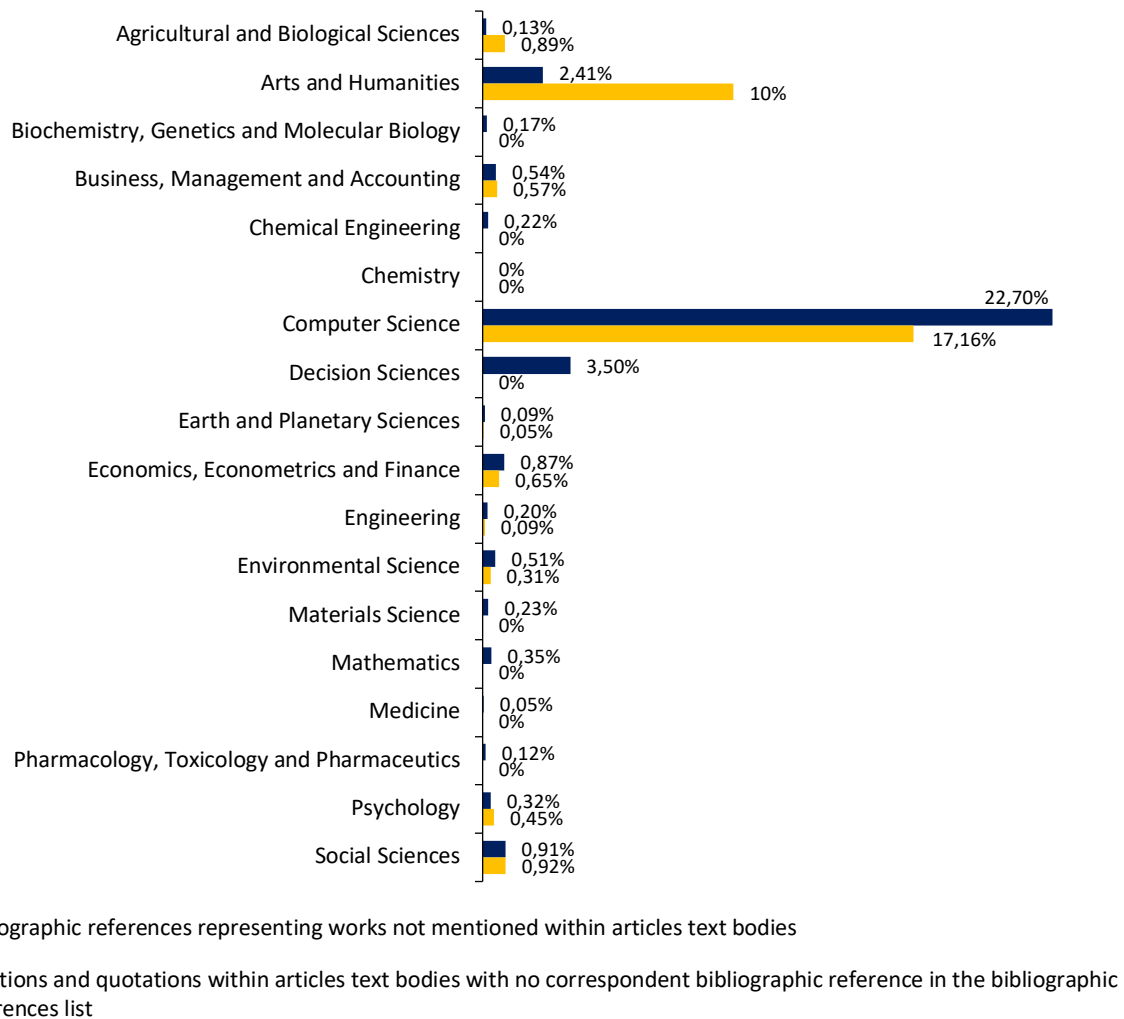

■ Bibliographic references representing works not mentioned within articles text bodies

■ Mentions and quotations within articles text bodies with no correspondent bibliographic reference in the bibliographic references list

**Graphic 6. Percentual distribution of bibliographic references not denoted in text bodies and mentions and quotations not corresponding to a bibliographic reference included in the article's bibliographic reference list per subject area.**

It is expected that the number of bibliographic references found in a bibliographic references list is equivalent to the number of works cited by the corresponding citing work. However, this situation was different among the articles of our sample. Graphic 6 shows the rates of bibliographic references that were not denoted in the text body and vice-versa. The in-text reference pointers associated with mentions and quotations that did not match bibliographic references were counted as non-cited mentions or quotations. The subject areas for which we did not observe any of these behaviours – namely Dentistry, Energy, Health Professions, Immunology and Microbiology, Multidisciplinary, Neuroscience, Nursing, Physics and Astronomy and Veterinary – are not shown in Graphic 6. Also, it is worth mentioning that data extracted from articles of a single Computer Science journal are entirely responsible for the rates retrieved for this discipline shown in Graphic 6.

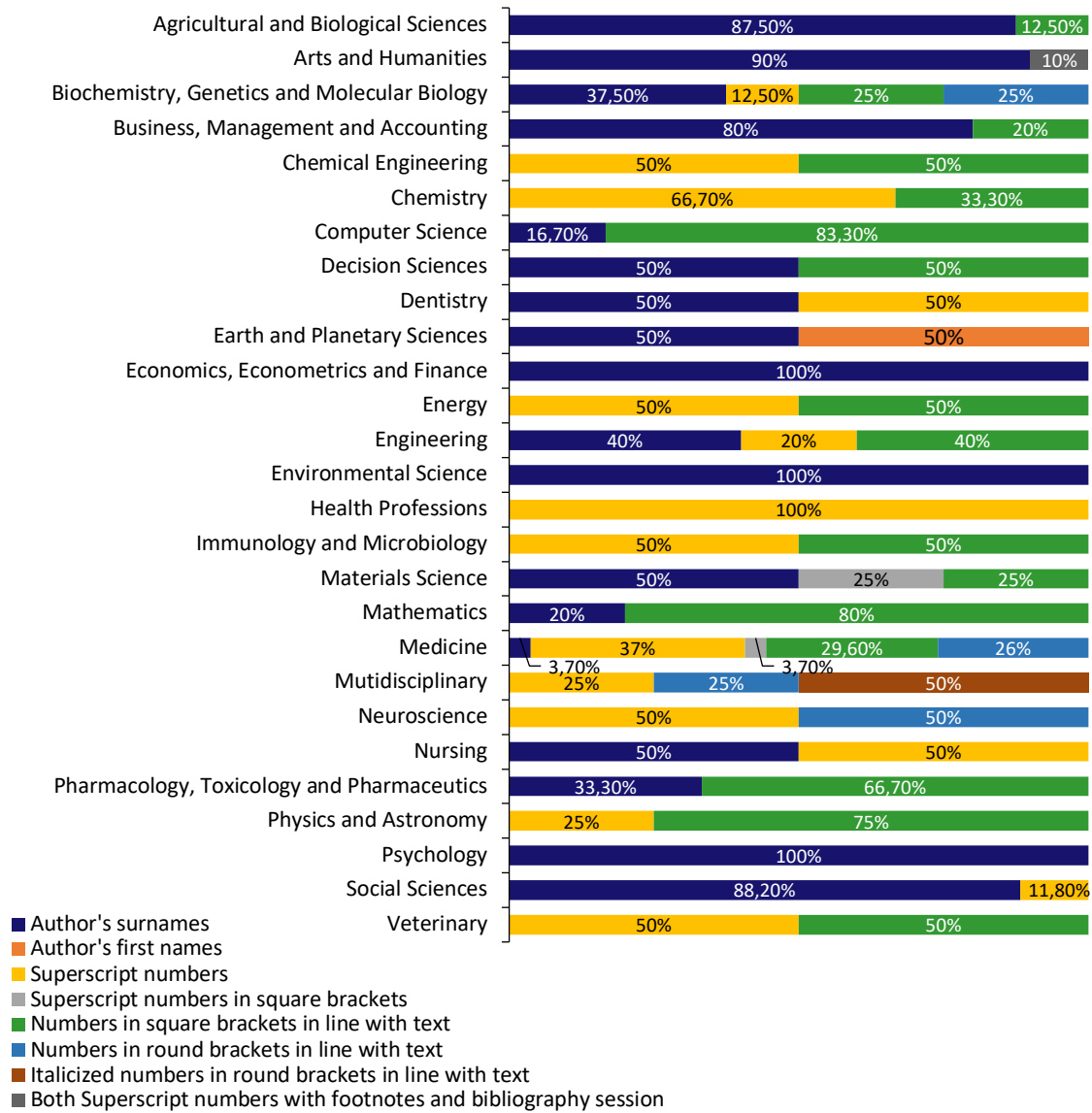

**Graphic 7. Percentual distribution of articles, according to the structure of in-text reference pointers across subject areas.**

Graphic 7 addresses the structure of in-text reference pointers across the disciplines. Except for a few subject areas – i.e., Economics, Econometrics and Finance, Environmental Science, Health professions, and Psychology – there is no uniformity in the way in-text reference pointers are used in text bodies, neither across subject areas nor across subject categories. Biochemistry, Genetics and Molecular Biology, Engineering, and Medicine are among the subject areas with the greatest variety of in-text reference pointers formats. According to Graphic 1, those disciplines are also among the ones adopting the widest variety of reference styles.

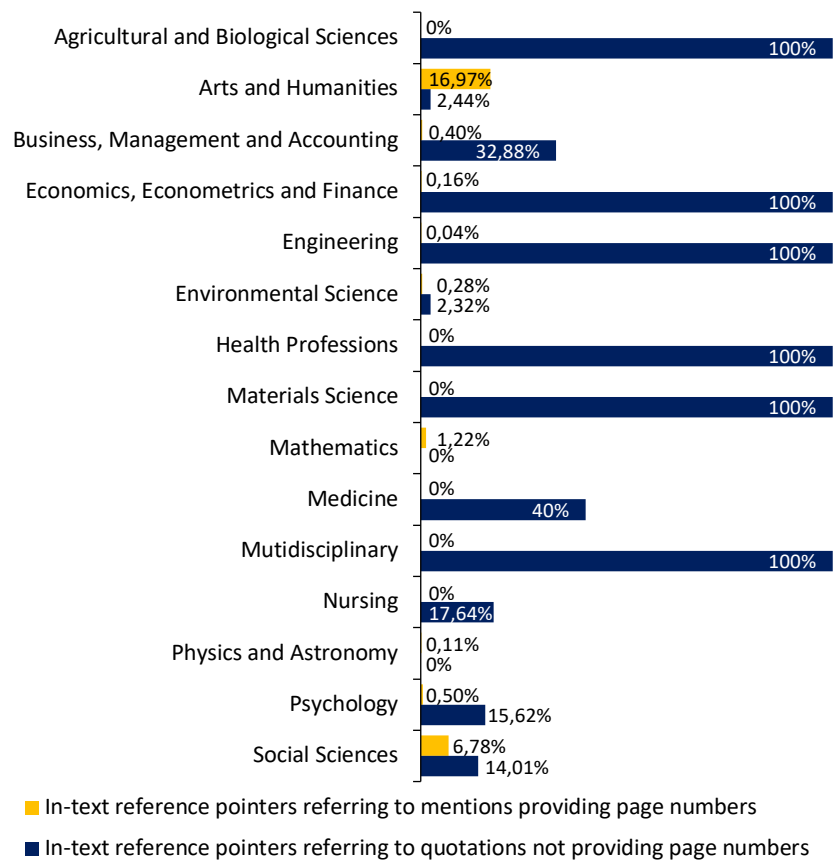

**Graphic 8. Percentual distribution of in-text reference pointers referring to mentions and quotations considering the provision of the page numbers of the cited passages in the cited works**

The provision of the page numbers in which the cited passage (Graphic 8) can be found within the cited work is usually considered by reference styles as an optional element for in-text reference pointers referring to mentions and a mandatory element for those referring to quotations. In all the articles in Biochemistry, Genetics and Molecular Biology, Chemical Engineering, Chemistry, Computer Science, Decision Science, Dentistry, Earth and Planetary Sciences, Energy, Immunology and Microbiology, Neuroscience, Pharmacology, Toxicology and Pharmaceutics, and Veterinary, we did not notice in-text reference pointers referring to mentions providing page numbers nor quotations not providing it within the articles from such disciplines.

|  | Run-in quotations | | | | | | | Long quotations | | | | | |
|---|---|---|---|---|---|---|---|---|---|---|---|---|---|
|  | 1 | 2 | 3 | 4 | 5 | 6 | 7 | A | B | C | D | E | F |
| **Agricultural and Biological Sciences** | 100% | 0% | 0% | 0% | 0% | 0% | 0% | 0% | 0% | 0% | 0% | 0% | 0% |
| **Arts and Humanities** | 47.06% | 0% | 2.94% | 41.18% | 0% | 0% | 0% | 32.35% | 2.94% | 17.65% | 0% | 0% | 0% |
| **Biochemistry, Genetics and Molecular Biology** | 0% | 0% | 0% | 0% | 0% | 0% | 0% | 0% | 0% | 0% | 0% | 0% | 0% |
| **Business, Management and Accounting** | 73.33% | 0% | 0% | 6.67% | 6.67% | 0% | 0% | 26.67% | 0% | 0% | 0% | 0% | 0% |
| **Economics, Econometrics and Finance** | 100% | 0% | 0% | 0% | 0% | 0% | 0% | 0% | 0% | 0% | 0% | 0% | 0% |
| **Engineering** | 83.33% | 0% | 0% | 16.67% | 0% | 0% | 0% | 0% | 0% | 0% | 0% | 0% | 0% |
| **Environmental Science** | 40% | 0% | 0% | 40% | 0% | 0% | 0% | 40% | 0% | 0% | 0% | 0% | 0% |
| **Health Professions** | 100% | 0% | 0% | 0% | 0% | 0% | 0% | 0% | 0% | 0% | 0% | 0% | 0% |
| **Materials Science** | 100% | 0% | 0% | 0% | 0% | 0% | 0% | 0% | 0% | 0% | 0% | 0% | 0% |
| **Medicine** | 100% | 0% | 0% | 0% | 0% | 0% | 0% | 0% | 0% | 0% | 0% | 0% | 5.88% |
| **Multidisciplinary** | 100% | 0% | 0% | 0% | 0% | 0% | 0% | 0% | 0% | 0% | 0% | 0% | 0% |
| **Nursing** | 100% | 0% | 0% | 0% | 0% | 0% | 0% | 0% | 0% | 0% | 0% | 0% | 0% |
| **Physics and Astronomy** | 100% | 0% | 0% | 0% | 0% | 0% | 0% | 0% | 0% | 0% | 0% | 0% | 0% |
| **Psychology** | 90% | 0% | 0% | 0% | 0% | 0% | 10% | 0% | 0% | 0% | 0% | 0% | 0% |
| **Social Sciences** | 70.69% | 0% | 0% | 24.14% | 1.72% | 0% | 0% | 18.97% | 0% | 12.07% | 0% | 3.45% | 0% |

**Table 2. Percentual distribution of articles per subject area according to the markups adopted for run-in quotations and long quotations**

*Legend for Table 5*

| 1 | Double quotation marks |
|---|---|
| 2 | Double quotation marks and italic font |
| 3 | Double quotation marks some italics and some not italics |
| 4 | Single quotation marks |
| 5 | Part in single, part in double quotation marks |
| 6 | Italic normal font size |
| 7 | Part in double quotation marks, part in italic normal font size |

| A | Indented in a smaller font size |
|---|---|
| B | Indented smaller font size, part italics, part not italics |
| C | Indented in normal size font |
| D | Indented, in double quotation marks and italic normal font size |
| E | Indented, in double quotation marks and italic smaller font size |
| F | Aligned text in a smaller font size |

Table 2 introduces the variety of markups adopted by articles to identify the length of quotations within text bodies and to distinguish them from the self-authored content.

The criteria for determining the use of markups for long quotations differed. For instance, we noticed articles adopting indentation as markup for quotes longer than 80 words; others considered the length of the quotation (in lines) as a parameter for determining the use of indentation or not. Some articles have indented quotations longer than two lines, while others have indented quotations longer than four. In some cases (i.e., one journal in Social Science, one in Business and Accounting, and two in Arts and Humanities), single and double quotation marks were considered as markups for delimiting quotations within the same article. Another point observed is that there is no uniformity even among the unusual scenarios. For instance, some in-text reference pointers concerning mentions in the same article provide the pagination of the mentioned excerpt in the cited work, and others do not.

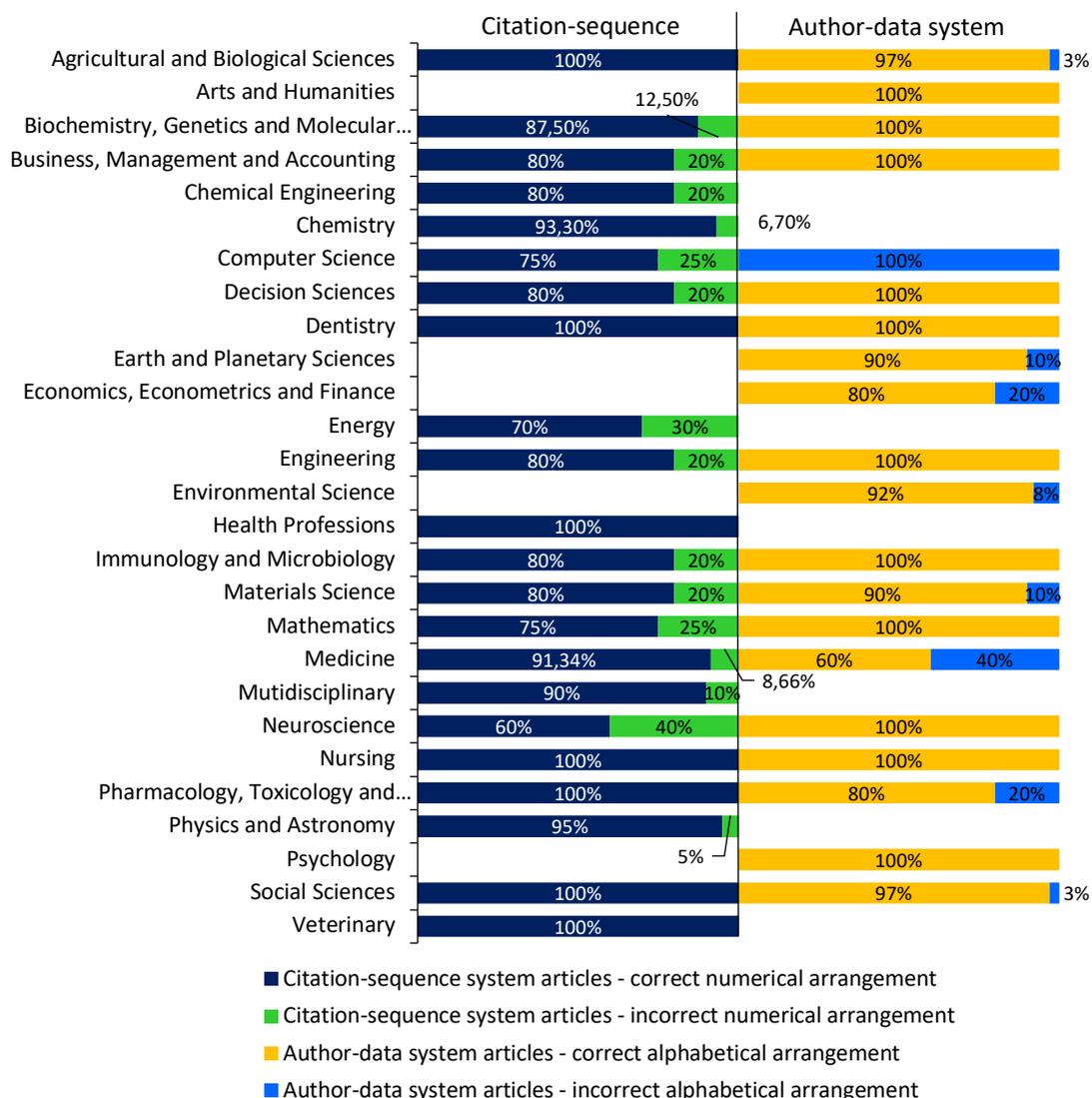

**Graphic 9. Percentual distribution of articles per subject area, considering the sorting of bibliographic references within the bibliographic references list**

Numbers accompanying the bibliographic references in the bibliographic reference lists do not always correspond to the format of those denoting them through in-text reference pointers in the text body, nor are they properly assorted. Graphic 9 indicates that, on average, 10.48% and 7.93% of articles adopting the citation-sequence system and author-data system, respectively, do not provide a proper numerical/alphabetical sorting in their references list.

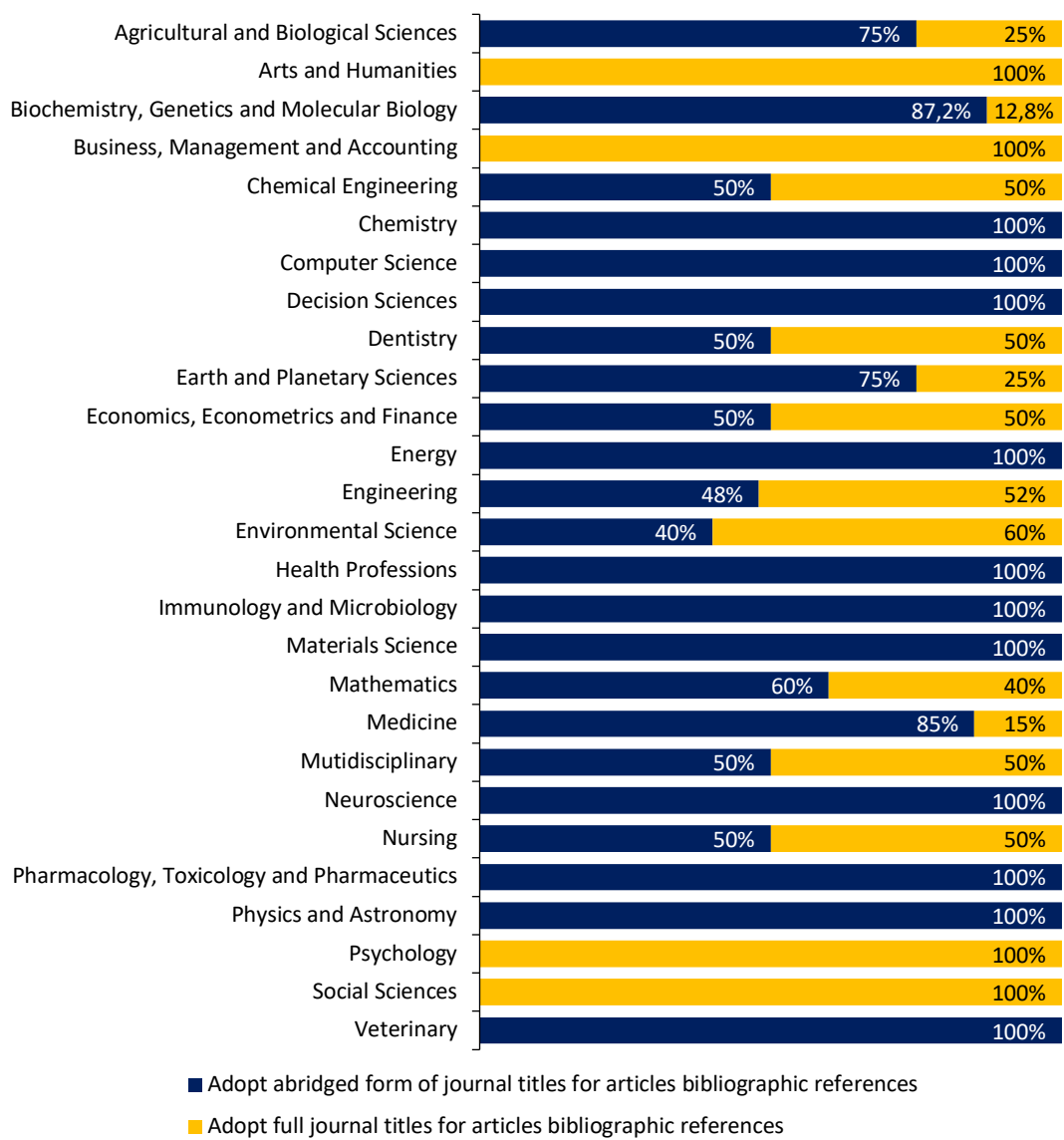

**Graphic 10. Percentual distribution of journals according to the format in which cited journals' titles are transcribed in bibliographic references: in full or in abridged format.**

Concerning the titles of cited journals in bibliographic references, the abridged format is adopted by 67.41% of journals. On average, Health Sciences, Life Sciences, and Physical Sciences adopt abridged versions of journals' titles, while Social Sciences journals provide journals' titles in full. Only 14.81% of disciplines consider only the full title of articles: Arts and Humanities, Business, Management and Accounting, Psychology and Social Sciences, as shown in Graphic 10. Graphic 11 provides a complementary view of the abridged journal titles, highlighting the source used to define the abbreviations for the cited journal. In 65.22% of the subject areas, different abbreviation sources for journal titles were detected across journals of the same area. However, 46.23% of journals in our sample did not specify any source.

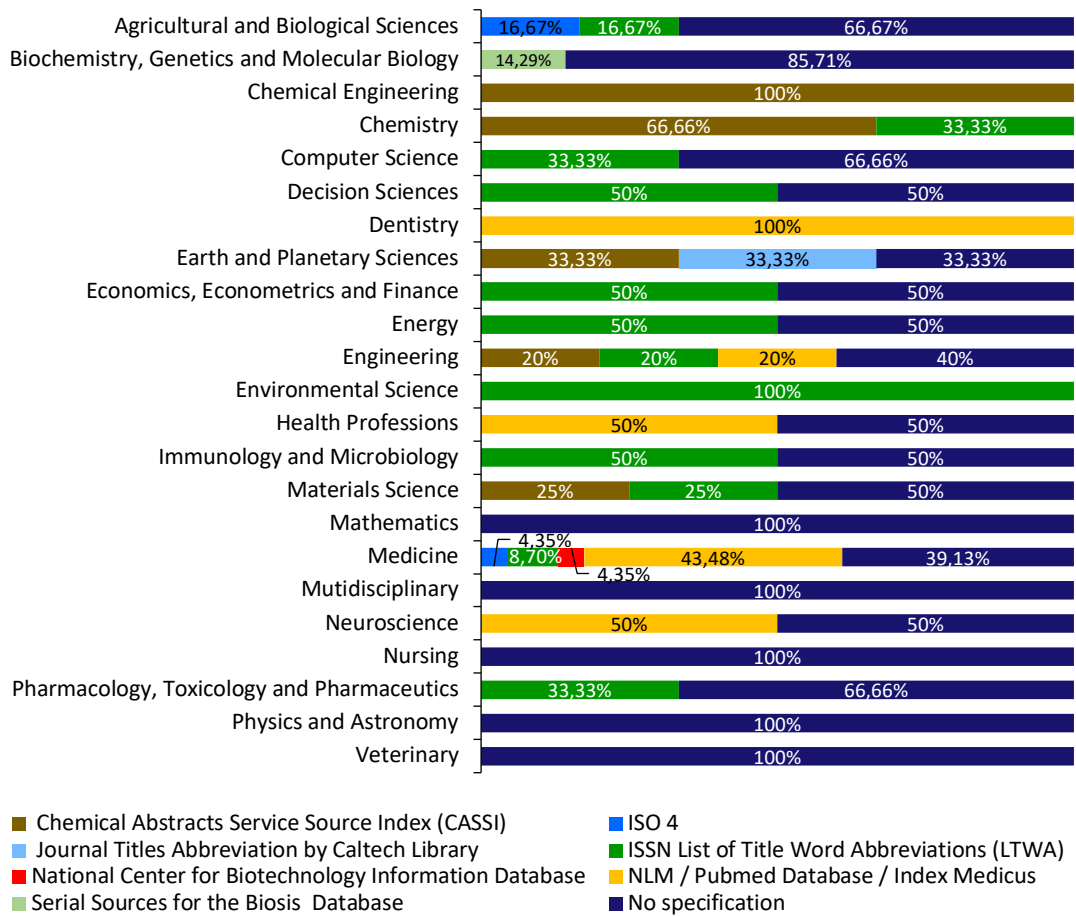

**Graphic 11. Percentual distribution of journals considering the source adopted for abbreviations of journal titles of cited articles within bibliographic references**

| Subject Areas \ Reference managers | Products supporting CSL Styles | Bibstyle[4] | Bibtex[5] | Endnote | Mendeley | Reference Manager | Zotero | Any reference manager software | Not identifiable |
|---|---|---|---|---|---|---|---|---|---|
| Medicine | 0% | 0% | 0% | 90% | 27% | 0% | 18% | 0% | 0% |
| Nursing | 0% | 0% | 0% | 100% | 0% | 0% | 0% | 0% | 0% |
| Veterinary | 50% | 0% | 0% | 50% | 50% | 0% | 0% | 0% | 0% |
| Dentistry | 0% | 0% | 0% | 100% | 0% | 0% | 0% | 0% | 0% |
| Health Professions | 0% | 0% | 0% | 100% | 0% | 0% | 0% | 0% | 0% |
| Arts and Humanities | 0% | 0% | 0% | 100% | 0% | 0% | 0% | 0% | 0% |
| Business, Management and Accounting | 50% | 0% | 0% | 50% | 50% | 0% | 0% | 0% | 0% |
| Decision Sciences | 50% | 0% | 0% | 50% | 50% | 0% | 0% | 0% | 0% |
| Economics, Econometrics and Finance | 50% | 0% | 0% | 0% | 50% | 0% | 0% | 50% | 0% |
| Psychology | 50% | 0% | 0% | 50% | 50% | 0% | 0% | 0% | 0% |
| Social Sciences | 0% | 0% | 0% | 100% | 0% | 0% | 20% | 0% | 0% |
| Agricultural and Biological Sciences | 20% | 0% | 0% | 80% | 20% | 20% | 20% | 0% | 0% |
| Biochemistry, Genetics and Molecular Biology | 0% | 0% | 0% | 100% | 50% | 0% | 50% | 0% | 0% |
| Immunology and Microbiology | 50% | 0% | 0% | 0% | 50% | 0% | 0% | 0% | 0% |
| Neuroscience | 0% | 100% | 0% | 100% | 0% | 0% | 0% | 0% | 0% |
| Pharmacology, Toxicology and Pharmaceutics | 50% | 0% | 0% | 50% | 0% | 50% | 50% | 0% | 0% |
| Chemical Engineering | 0% | 0% | 0% | 100% | 0% | 0% | 0% | 0% | 0% |
| Chemistry | 33% | 0% | 0% | 67% | 33% | 0% | 0% | 0% | 0% |
| Computer Science | 33% | 0% | 33% | 33% | 33% | 0% | 0% | 0% | 0% |
| Earth and Planetary Sciences | 25% | 0% | 25% | 25% | 25% | 0% | 0% | 0% | 0% |
| Energy | 100% | 0% | 0% | 0% | 100% | 0% | 0% | 0% | 0% |
| Engineering | 0% | 0% | 0% | 50% | 50% | 0% | 0% | 0% | 0% |
| Environmental Science | 33% | 0% | 0% | 67% | 33% | 0% | 0% | 0% | 0% |
| Materials Science | 50% | 0% | 0% | 0% | 50% | 0% | 0% | 0% | 50% |
| Mathematics | 33% | 0% | 33% | 33% | 33% | 0% | 0% | 0% | 0% |
| Physics and Astronomy | 0% | 0% | 0% | 0% | 0% | 0% | 0% | 0% | 0% |
| Multidisciplinary | 0% | 0% | 0% | 0% | 0% | 0% | 0% | 0% | 0% |

**Table 3. Percentual distribution of journals by subject areas recommending reference managers to authors.**

The lack of standardisation is also reflected even within the title of the bibliographic references section, in which 12 variations were detected across articles of our sample. On average, 93% of the articles entitle their bibliographic references session as "References". Most publishers (53.01%) suggest authors use reference managers to format bibliographic references of their works. Table 3 shows each reference manager's recommendation rate per subject area. Some journals simultaneously recommend multiple reference styles, some recommend none, and others refrain from mentioning a particular one (i.e., "any reference manager" column), although they recommend the use of reference managers.

---

[4] Bibstyle is a reference style adopted by some academic journals.
[5] BibTeX is a fomatting tool used in LaTeX documents. And LaTeX is a marking system or program for publishing high quality typographic documents, specific for the elaboration of scientific texts.

Journals also can provide tools for exporting citation metadata, formatted according to specific reference styles or reference standards guidelines (e.g., Chicago and ABNT), simple text, HTML or MS Word file (with no specification of the reference style in which is formatted), machine-readable files (i.e. Endnote or Zotero files) or, interchangeable files formats (e.g. CSV and RIS, which are usually readable by reference managers like Endnote, Mendeley and Zotero). It should be mentioned that publishers refer to structured formats (e.g. machine-readable, like RIS) and unstructured formats (e.g. plain text) as the same things. Table 4 highlights the most provided formats in which bibliographic references are available.

| Kind of format | Reference styles or formats provided to export bibliographic references | Percentual average of journals per subject area |
|---|---|---|
| **Structured format (machine-readable)** | Bibtex | 69.9% |
| | RIS | 68% |
| | Refworks | 56% |
| | Endnote | 41% |
| | Mendeley | 28.6% |
| | Medlars | 24.2% |
| | Medline / Pubmed | 4.9% |
| **Unstructured format (plain text)** | Text (a.k.a. simple text or plain text) – refer to journals providing bibliographic references with no specification of the reference system in which it is formatted | 60.3% |
| | Chicago | 4% |

Table 4. Exportation files and files formats adopted by publishers for exporting articles' bibliographic metadata

The core objective of systematically structuring bibliographic metadata as in-text reference pointers and bibliographic references is to assure the identification of a cited content, i.e., a work or a passage, by providing (at least) the minimum elements supporting the reader in seeking tasks within a bibliographic catalogue. Of course, such minimum elements may change depending on the particular type of the publication. For instance, a bibliographic reference to a journal article should include at least the authors, title, year of publication, journal name, volume, issue, and pagination. Instead, a reference to a born-digital publication, such as a dataset, should include authors, the year of publication, the version number, and the repository where it has been deposited. In addition, in the case of quotations, the specific pagination of the passage quoted should also be included in the in-text reference pointer denoting the bibliographic reference of a cited publication[6]. Graphic 12 shows the percentual distribution of in-text reference pointers and bibliographic reference pairs that correctly match the work's identification goals and, consequently, allow easy access. In this context, "easy access" is assumed as the set of descriptive elements addressed by the in-text reference pointers and bibliographic references sets properly supporting the identification of the work containing the quoted passage and their location within the cited work, without the need to proceed complementary seeking in indexes or summaries, or to scroll down an electronic content, or leaf through a printed publication to get to the exactly cited passage. We noticed that on average, 70% of articles having quotations do not provide easy access to the cited content.

---

[6] See (Santos et al., 2023) for a more in-depth analysis of the basic metadata used in bibliographic references referring to different types of publications.

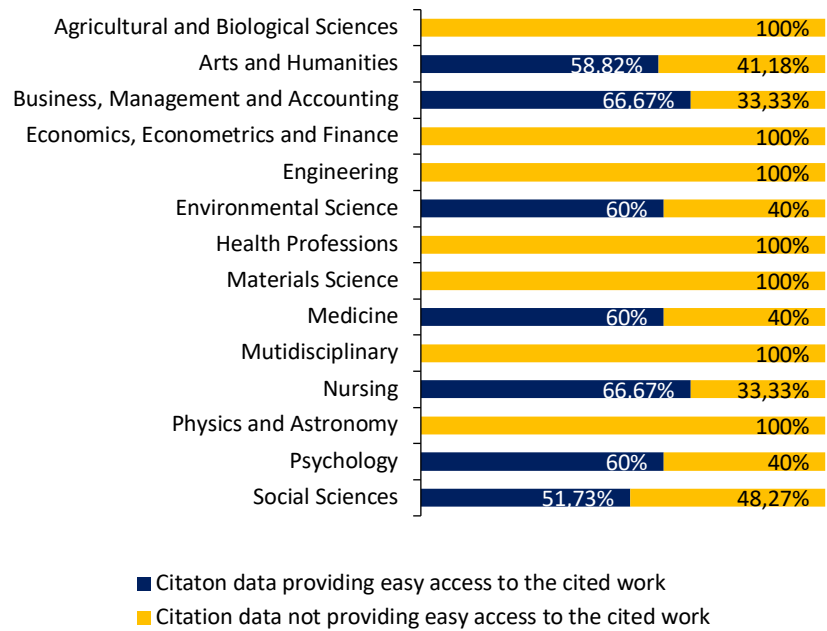

**Graphic 12. Percentual distribution of articles (considering only disciplines with articles with quotations) according to the facility of accessing the cited work, considering data provided by in-text reference pointers concerning quotations and the related bibliographic references**

## 5. Discussion

### 5.1. Same data, several representations

Our analysis showed that each subject area adopts different styles within their periodical publications, as also observed by Tovaruela Carrión et al. (2017), and that there is a variation in the adoption of reference styles even among journals from the same discipline. For instance, we detected 11 different reference styles among journals from the Medicine subject area (Graphic 1).

The percentage of journals adopting their "own reference styles" surpassed all those concerning other reference styles. Indeed, 100% of Multidisciplinary journals adopt their "own reference styles". Disciplines adopting "own reference style" should be considered the less engaged with the normalisation matters since the lack of standardisation is one of the main contributing factors for citation errors (Sweetland, 1989). Besides, customisation is expensive (Hoffman 2009, p. 636) and disagrees with standardisation purposes since a customised version of a reference style configures a new reference style.

The wide variety of reference styles supports Tovaruela Carrión et al.'s (2017) arguments that the presentation of bibliographic references is one of the most frequent problems in scientific literature. In addition, the multiple coexisting editions of some reference styles, i.e., AMA, APA, and Chicago, represent issues in the normalisation field. Journals must adopt one of those multiple versions of the same reference style and specify the precise edition of the adopted reference style. Nevertheless, our analysis suggested the opposite behaviour in several cases. E.g., while 25.93% of Medicine journals adopt the AMA reference style, 7.41% of journals indicate the AMA $9^{th}$ edition, 11.11% indicate the AMA $10^{th}$ edition, and 7.41% do not provide accurate instructions concerning which particular edition of such reference style authors should consider.

### 5.2. The disadvantages of reinventing the wheel

Analysing Graphic 5, we can speculate about a direct relation between journals adopting reference styles authored by their respective publishers and the rate of reference styles not providing clear formatting guidelines for citing and referencing metadata compared with those widely adopted ones. In particular, we observed a need for more details on how to reference specific types of publications, such as online grey literature and, often, poor instructions on referencing and citing practices, thus resulting in errors. For instance, Chemical Engineering, Neuroscience, and Nursing subject areas were the most accurate disciplines in this context, providing clear and complete guidelines. Indeed, none of them adopts "own reference styles" but rather well-known and shared guidelines. In addition, the elaboration of yet another reference style configures it as a duplicated work since its descriptive content is usually limited, and its purposes could be addressed more appropriately by the well-known and shared reference styles. As Sweetland (1989) claimed, publishers could be more zealous on behalf of standardisation of bibliographic metadata matters.

From the data gathered, we also noticed a need for more commitment from some publishers to standardisation issues when they instruct authors on how to proceed with citing and referencing bibliographic metadata. We encountered situations where the link to access the reference style adopted by a journal did not work. Some publishers do not even provide information concerning the reference style adopted by their journals within the instructions to authors made available on their web pages – not to mention the case of a particular journal recommending authors to consult the bibliographic references lists of the articles published in the previous issues of the journal and consider them as a model for formatting their bibliographic references. In these cases, publishers risk reiterating errors introduced in previous journal issues (Sweetland, 1989). In addition, authors may need to cite different types of works than those referenced in the sample articles, and they have to proceed with such a formatting task using their intuition. Considering all these issues, "journal editors and referees could pay greater attention to the quality and quantity of references" (Cronin, 1982).

### 5.3. The citation systems: citation-sequence and author-date

Graphic 4 evinces a balanced scenario between journals adopting the author-date citation system and the citation-sequence system. However, it is worth highlighting the remaining 0.2% of the articles (corresponding to 5.9% of Social Science articles adopting The Chicago Manual of Style, 16$^{th}$ ed.), adopting both author-date and citation-sequence systems, within the same articles. In those cases, mentions and quotations are marked up with superscript numbers, denoting the bibliographic reference of the cited work, which is provided in a footnote (citation-sequence system). Simultaneously, the bibliographic reference lists consider the same bibliographic references addressed in the footnotes in an alphabetical assortment (author-date system). This scenario denotes unpropped behaviour of the publisher regarding citation matters which goes against standardisation purposes on facilitating the correlation between bibliographic metadata and the works they address.

Considering the thousands of existing reference styles, it can be appropriate to present bibliographic metadata differently if they comply with a bibliographic style. Indeed, one contributing factor to the vastness of the citation styles is the fact that certain styles have variations on their citing systems, like the Chicago style (Barbeau, 2018) and, especially in these cases, publishers should devote efforts to explaining to authors the interpretation and selection of the alternatives offered by the reference style, since doing customisations and amendments to such styles do not make the normalising tasks simpler. It is worth mentioning that Social Sciences was the subject area which showed the second highest rates of reference styles adopted and the only discipline showing an article using both citation systems, which supports the claim that the excesses in the variety of guidelines and the omission of editors can be disruptive to standardisation issues.

### 5.4. The correspondence between cited works and bibliographic references

Patiño Díaz (2005, p. 21, our translation) defines a bibliographic reference as "the data that indicate to the reader whose **quote** he is reading and where to find it in its original version", i.e., the cited work. Masic (2013, p. 150) complements that "in scientific circles, the reference is the information necessary to the reader in identifying and finding **used sources**". However, in addition to the "identifying function", some bibliographic references providing hypertext links or DOI hyperlinks also accomplish the "finding function" referred to by Masic (2013), although it is primarily up to the library catalogues. At this point, we must complement Patiño Díaz's statement (2005): bibliographic references suggest not only the identification of the sources of quotes but also the sources on which mentions are based.

However, bibliographic references referring to works not cited along the text body and the reverse situation, i.e., mentions and quotations without a corresponding bibliographic reference in the bibliographic reference list, were identified in our sample. In such cases, the reader might be prevented from retrieving the cited content, configuring a contradiction to the five Ranganathan's Laws (Zabel & Rimland, 2007).

Providing data that favours online access to cited works may be considered a courtesy but an efficient way to facilitate the identification and access of cited works. However, it should be noted that reference styles rarely provide clear and enough instructions on this matter, favouring different interpretations for similar approaches within different reference styles.

### 5.5. Structuring in-text reference pointers structuring

In-text-reference pointers are also subject to the ravages of the multiplicity of reference styles and the superficiality of formatting instructions. We observed no standardisation, neither on metadata addressed within in-text reference pointers referring to mentions and quotations nor on their formatting instructions. Considering the several ways in-text reference pointers appear in text bodies, readers may confuse them with other elements. For instance, mathematical and chemical formulas frequently use superscript numerical characters, which can easily be confused with in-text reference pointers since both can share the same format. In addition, we noticed that, in most disciplines, there is not an established standard behaviour on how to present in-text reference pointers, even considering the choices between author-date and citation-sequence styles, which are the main categories that guide the definition of the in-text reference pointer structure.

For instance, one of the long-indented quotations in our sample reproducing a passage from Jamie Dreier (Kurth, 2019) is concluded by the following in-text reference pointer: "(2014a: 178; also: Korsgaard 2008; Gibbard 1990)". This example suggests misunderstandings regarding mentions and quotations and the proper way of denoting them within works' text bodies by using in-text reference pointers. Since a quotation is a literal and exact transcription of one or more passages from a cited work into a citing work, it is impossible to cite more than one document per quoted passage simultaneously. This fact supports the claim that, besides the epistemological issues pointed out by Galvão (1998), such conceptual unclearness that hangs over Information Science also has a practical effect on identifying certain elements in scientific works.

### 5.6. Page numbers provision in in-text reference pointers

The provision of the pagination within the in-text reference pointers referring to mentions and quotations, where the cited content can be found in the cited work, is not uniform since most reference styles do not provide instructions for addressing this aspect. Indeed, we did not notice a common habit among our sample articles. For instance, as discussed in Graphic 8, the pagination metadata are sometimes provided when considered an optional element, i.e., in-text reference pointers referring to mentions. It is not always offered when it is regarded as a mandatory element, i.e., in-text reference pointers referring to quotations, which makes the task of seeking the original text in the cited work harder.

### 5.7. The transcription of journal titles in bibliographic references

Our analysis showed that, in most journals from Health Sciences, Life Sciences, and Physical Sciences, the titles of the journals in which cited articles were published are provided in the abridged format. The main reason may be making bibliographic references shorter. However, in such cases, it may be difficult for a reader to precisely interpret to which journal such an abridged title refers. For instance, according to the ISO 4 standard, the journal title "European Physical Journal" should be abbreviated as "Eur. Phys. J.". Considering such abbreviation from the perspective of the reader, who is not supposed to know ISO 4 guidelines, the abbreviation "Phys" may be interpreted as "Physics", "Physical" or "Physician". Those misinterpretations may represent difficulties in retrieving the correct original journal title, mainly because the source from where such abbreviations are taken is not provided within articles but only within the instructions for authors, whenever provided at all.

We noticed that 46.2% of journals adopting abridged journal titles, on average, do not provide the source on which such abbreviations should be based, not even to authors and, within the remaining sample, we detected seven different recommended sources for journal titles abbreviations. Besides favouring different abbreviations for the same journal title, such a range of sources goes against the principles of standardisation. Even considering the possibility of making bibliographic references shorter, abridging journal titles may be viewed as a non-sense practice in the era of the electronic universe, especially considering IFLA-LRM approaches on the "data and functionality required by end-users (and intermediaries working on behalf of end-users) to meet their information needs" (Riva, Le Bœuf & Žumer, 2017, p. 15).

### 5.8. The use of reference managers for managing bibliographic metadata

According to data shown in Table 3, publishers usually recommend authors in using reference managers, like Endnote or Mendeley, for dealing with bibliographic metadata. In theory, reference managers can solve standardisation problems within bibliographic references. In practice, using such tools (should) demand careful monitoring of reference styles. Whenever an update or amendment is identified, such data should be immediately introduced within the reference manager's stylesheets. We got the impression that publishers commit themselves to provide tools for the automatic writing of bibliographic references rather than offering instructional resources on such issues, which seems to confirm Sweetland's (1989) statements that the lack of training in the norms and purposes of the bibliographic citation.

### 5.9. Exporting citations tools within publisher's webpages

Table 4 shows that publishers may provide tools for exporting bibliographic metadata of the articles they publish within their journal's web pages. However, different output formats are presented as similar. For instance, some publishers provide bibliographic metadata in human-readable textual formats, for which usually there is no indication of which reference style it is formatted to. Since authors may copy and paste such textual bibliographic references into their own bibliographic references lists, by collecting such data from several publishers, they might have bibliographic references formatted under different criteria, i.e., different reference styles' guidelines, mixed in the same bibliographic reference list.

### 5.10. Does bibliographic metadata facilitate access to scientific information?

The rate of bibliographic references whose metadata was not enough to clearly identify the work they reference is a matter of concern. It suggests that reference styles usually do not provide clear and comprehensive instructions on presenting bibliographic metadata. Consequently, such guidelines are not adequately understood by the authors, who end up providing bibliographic metadata in the way they (erroneously) believe it should be the most proper one. In this perspective, publishers' passive stance on the issues concerning citing and referencing description and normalisation confer them a significant role in compounding the counterproductive scenario on bibliographic normalisation matters, which might at least be less problematic from the authors' and readers' perspectives.

Álvarez De Toledo (2012) states that scientific styles are methods of writing, structuring, representing, and organising scientific content, including mentions, quotations, and bibliographic references. Therefore, citing and referencing habits depend on the guidelines of the adopted scientific style, i.e., there is no universal habit in this scenery. Taylor (2006) states that the task of cataloguing primarily is to develop and apply standards to create bibliographic records that describe and provide access to information packages. Such statements reinforce the understanding that bibliographic catalogues complement the functions of bibliographic references as information access facilitators. Substituting *cataloguing* in Taylor's notion with *bibliographic reference standardisation* makes her statement still valid. The principle of user convenience assumes that cataloguers can objectively determine the user's needs and will know how to customise bibliographic records to meet these needs (Hoffman, 2009). However, it should be noted that multidisciplinary is becoming increasingly necessary and not only an added value (Martins, 2007). Facing this, it can be assumed that, in contrast to cataloguing, citing and referencing matters should not consider the *local user's* but the *global user's* concepts and needs. For instance, there can be no assurance that a medical work will not support a Social Science work. This weakens the statement that the existence of multiple reference styles is needed to fulfil the specific needs of each discipline. Indeed, the multidisciplinary point of view invalidates the local user's point of view and either enlarges or enhances the target audience of a particular work to the whole scientific community. Second, the users and their real needs become unclear in this context. Within the bibliographic metadata standardisation domain, there are no local users; therefore, the customisation of bibliographic references loses its sense. In such a context, questions raised by Hoffman (2009) referring to the cataloguing universe, i.e. "how can local users' needs be met?" or "who is responsible for meeting users' needs in cataloging?" and "what is the "right" way(s) for cataloging to help users and ensure equitable access to materials?" do not apply to citing and referencing world. So, the widely adopted reference styles, such as Vancouver and Chicago, would suffice the scientific community's needs and expectations on bibliographic matters. Bibliographic references do not address problems and needs but provide metadata that allows one to clearly identify a particular work and seek it within bibliographic catalogues.

## 6. Conclusion

Reference styles do not provide broad and clear coverage of all aspects concerning bibliographic metadata description in the form of bibliographic references. Consequently, several aspects of the scientific universe evinced a lack of standardisation. In addition, not following any reference style or not following the adopted reference style properly suggests problems in the editorial process quality. Issues regarding the presentation of bibliographic references head the list of the most frequent problems in the scientific literature (Tovaruela Carrión et al., 2017). Authors and the journals where they publish are the most affected entities by the errors in citing and referencing (Ruiz-Pérez, Delgado López-Cózar & Jiménez Contreras, 2006), and this makes such a subject an object of investigation (Osca-Lluch et al., 2009). In this study, by examining several journals of a large set of scholarly disciplines, we considered three research questions related to a prior study by Sweetland.

As an answer to RQ 1, our data showed that the errors pointed out by Sweetland (1989) still hold today, plus other issues identified in the study. The way information is produced, stored, retrieved, used, and represented has changed since the time in which Sweetland developed his study. However, how information is approached by descriptive representation from the citing and referencing perspectives has remained the same. Besides, nowadays, authors can count on technological resources to help with citing and referencing metadata, i.e., reference managers. However, consciously or not, the increasing amount of reference styles contributes to multiple (and sometimes totally different) interpretations of similar guidelines, which ends up acting as a barrier to the standardisation of bibliographic metadata concerning citing and referencing matters. In the future, it would be interesting to extend the current corpus of documents we used in this analysis by considering articles published more recently (e.g. 2023) to see if the additional citing and referencing practices introduced recently, such as (Starr et al., 2015; Smith et al., 2016), may introduce an improvement at least for specific types of cited publications.

Referring to RQ 2, we noticed that the problems with citing and referencing metadata have increased since Sweetland's time. Publishers seem to consider mentions, quotations, and in-text reference pointers as separate elements when they are part of the same whole that complement each other. For instance, the adoption of technological resources, i.e. the reference managers, which intended to assist authors in dealing with such metadata, contributes to the non-accomplishment of standardisation purposes, by considering different versions of the same reference style and by not clearly providing such information.

On this matter, Masic (2013, p. 150) suggest that the "basic rule when listing the sources used is that references must be accurate, complete and should be consistently applied". However, since we do not know how readers will seek the information, we need more and more elaborated and practical systems and, above all, with the most international recognition (Patiño Díaz, 2005, p. 16). Citing and referencing should, therefore, be thought of and redesigned jointly with the descriptive representation revision to ensure a unique metadata language among the various instruments and contexts in which information is represented.

One last point to consider is the role of publishers in this scene. Publishers represent one of the most important means for scientific communications, based on a two-way relationship with authors. Publishers depend, at first, on the author's scientific production to have what to publish, while authors depend (not exclusively) on publishers to give publicity and visibility to their writings. This suggests that publishing is not only a way of making money, but it is also a mechanism to enhance scientific communication. Starting from this, some features attributed to the scientific text can be considered as ways to boost and facilitate the flow of scientific communication, as it favours both the connection between citing and cited works (and, consequently, the citations network) and between readers/researchers and works of interest, as Ranganathan's laws presuppose.

This study raises questions about whether the current citing and referencing practices meet users' needs. Indeed, this was one of the main reasons which substantiated the first steps in the revision of representative description, culminating in the development of FRBR. Our study provides the first insight into these matters across several disciplines. Still, its outcomes should not be considered definitive, and more in-depth discussions should be carried out. For instance, some errors may be more critical than others. In the future, we could extend the current study by rating the errors and the combination of errors, thus obtaining a more in-depth analysis of in-text reference pointers and bibliographic references. In addition, in the future, we can also extend the present analysis to measure, using some formal approach, correlations and causations to explain possible causes for the errors identified.

## 7. ACKNOWLEDGEMENTS

The authors would like to thank the Library Service of the School of Communication and Arts (ECA) of the University of São Paulo and the Library of the Department of Classical Philology and Italian Studies of the University of Bologna for their contribution to provide part of the articles that composed the sample of this study.

## 8. FUNDING SOURCES

The work of Silvio Peroni has been partially funded by the European Union's Horizon 2020 research and innovation program under grant agreement No 101017452 (OpenAIRE-Nexus) and the European Union's Horizon Europe research and innovation program under grant agreement No 101095129 (GraspOS).